\documentclass[runningheads]{llncs}
\usepackage[T1]{fontenc}
\usepackage{graphicx,verbatim}
\usepackage{hyperref} 

\usepackage{amsmath}

\usepackage{pgfplots}
\pgfplotsset{compat=1.18}
\usepackage{pgfplotstable}
\usepackage{tikz}
\usetikzlibrary{patterns, arrows.meta, positioning, calc}

\definecolor{clrPosterior}{RGB}{31,119,180}   
\definecolor{clrLateral}{RGB}{255,127,14}      
\definecolor{clrAnterior}{RGB}{44,160,44}      
\definecolor{clrAllViews}{RGB}{214,39,40}      
\definecolor{clrBaseline}{RGB}{148,103,189}    
\definecolor{clrGray}{RGB}{127,127,127}

\usepackage{pgfplots}
\pgfplotsset{compat=1.18}
\usepackage{enumitem}
\setlist{nosep, leftmargin=14pt}

\usepackage{mwe} 
\usepackage{subfig}
\usepackage{xcolor}

\begin{document}
\title{Prognostic Value of Lung Ultrasound Biomarkers for Readmission Risk in Congestive Heart Failure: A Pilot Data-Driven Analysis}
\titlerunning{Prognostic Value of Lung Ultrasound Biomarkers}


\author{Jana Armouti\inst{1} \and
Laura Hutchins\inst{2} \and
Jacob Duplantis\inst{2} \and
Thomas Deiss\inst{2} \and
Thales Nogueira Gomes\inst{2} \and
Keyur H. Patel\inst{2} \and
Seema Walvekar\inst{2} \and
Shane Guillory\inst{2} \and
Thomas H. Fox\inst{2} \and
Amita Krishnan\inst{2} \and
Ricardo Rodriguez\inst{3} \and
Bennett DeBoisblanc\inst{2} \and
Deva Ramanan\inst{1} \and
John Galeotti\inst{1} \and
Gautam Gare\inst{1}}

\authorrunning{J. Armouti et al.}

\institute{
Carnegie Mellon University, Pittsburgh, USA \and
LSUHSC Internal Medicine, New Orleans, USA \and
Cosmetic Surgery Facility LLC, Baltimore, MD 21093, USA
}
  
\maketitle              
\begin{abstract}
Hospital readmission within 30 days of discharge is a leading driver of morbidity, mortality, and avoidable healthcare expenditure in congestive heart failure (CHF). \cite{Harvey2021P004.Failure} Current clinical risk stratification tools rely primarily on non-imaging data and exhibit limited predictive performance. Point-of-care lung ultrasound (LUS) offers a sensitive, noninvasive window into the pulmonary congestion that characterizes CHF decompensation, yet its prognostic utility for readmission prediction remains largely unexplored.

We present a \textbf{pilot feasibility study}, the first systematic machine learning study using B-mode LUS acquired during hospitalization to predict 30-day CHF readmission. Quantitative spatiotemporal embeddings are extracted from a pretrained Temporal Shift Module (TSM)--ResNet-18 encoder, and interpretable biomarker features are separately evaluated. Through structured ablations over lung view, temporal representation, multi-view fusion, and cross-lung augmentation, we identify the key imaging factors driving readmission risk.

Our findings reveal that (1) dependent lower-lung regions (Left-3, Right-3) carry the strongest prognostic signal, consistent with their greater susceptibility to hydrostatic congestion; (2) temporal \emph{difference} features between sequential examinations substantially outperform single-timepoint representations, highlighting the importance of capturing disease trajectory; and (3) multi-view feature concatenation yields the best overall performance, with our top MLP model achieving an F1 score of \textbf{0.80} (95\% CI: 0.62--0.96). Biomarker analysis further reveals that pleural-line abnormalities, including breaks and indentations, are as informative as the canonical A-line and B-line markers. These results support POCUS-derived biomarkers as practical, interpretable tools for noninvasive CHF risk stratification.

\keywords{Prognosis \and Lung Ultrasound \and Longitudinal Modeling \and Biomarkers \and Interpretability \and Congestive Heart Failure}

\end{abstract}
\section{Introduction}
\label{sec:intro}

\begin{figure}[h]
    \centering
    \includegraphics[width=\columnwidth]{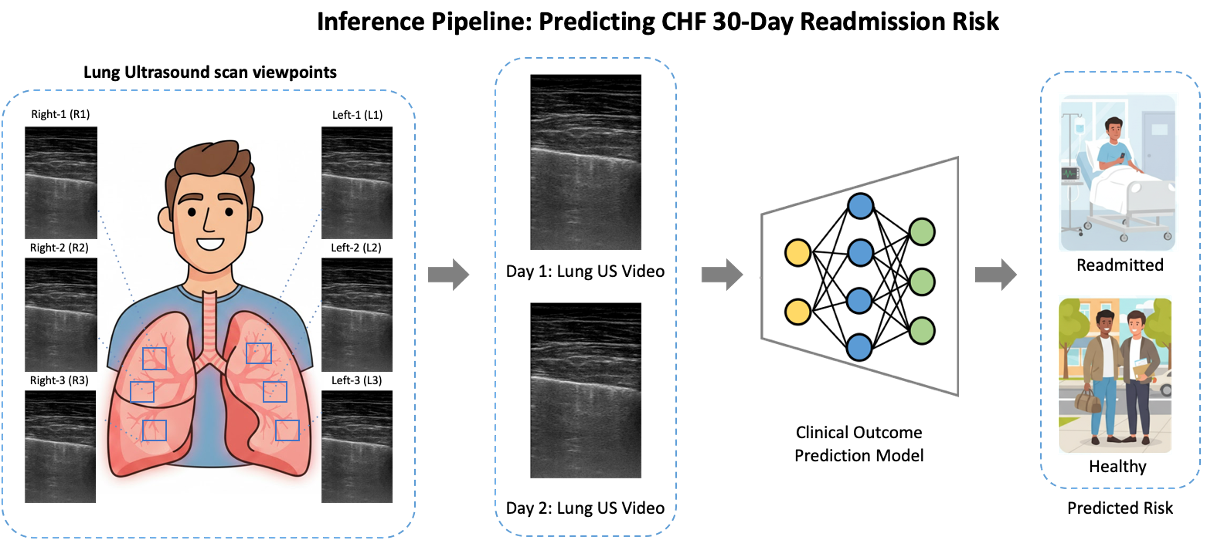}
    \caption{
        \textbf{CHF 30-day readmission prediction pipeline.}
        Six standardized LUS views (Right/Left 1--3, spanning upper anterior to dependent posterior regions) are acquired at two or more time points during the index hospitalization. B-mode video clips are encoded by a pretrained TSM--ResNet-18 to produce 512-dimensional spatiotemporal embeddings. Temporal difference features ($\Delta = \text{Day}_2 - \text{Day}_1$, capturing congestion trajectory) from all six views are concatenated into a single patient-level representation and fed to a downstream classifier to predict 30-day readmission risk.
    }
    \label{fig:overview_diagram}
\end{figure}

Hospital readmission in congestive heart failure (CHF) remains a major source of morbidity, mortality, and cost \cite{Sharma2022PredictingApproach,Harvey2021P004.Failure}. Predicting 30-day readmission is challenging, as pulmonary congestion evolves subtly and heterogeneously. Existing ML models based on electronic health records achieve only modest performance (AUC 0.60–0.72) \cite{Mortazavi2016AnalysisReadmissions,Sharma2022PredictingApproach}, partly because standard assessments poorly capture decompensation physiology.

CHF decompensation reflects fluid redistribution into the pulmonary interstitium and alveoli. Lung ultrasound (LUS) offers a real-time, noninvasive view of this process, quantifying A-lines (normal aeration), B-lines (interstitial fluid), and pleural abnormalities from B-mode clips \cite{gare2022learning,gare2022weakly}. Although established for diagnosing acute decompensation, its prognostic value for 30-day readmission remains underexplored.

Machine learning can model the subtle, spatial, and temporal patterns in LUS, integrating multiple lung zones and time points to capture congestion trajectories. Interpreting these models may also reveal clinically meaningful, underrecognized prognostic biomarkers.

\vspace{2em}
\textbf{Contributions.} We present the first structured pilot ML study of prognostic potential of point-of-care ultrasound (POCUS) for predicting 30-day CHF readmission. Using B-mode LUS scans collected during hospitalization and pre-discharge, we train ML models to estimate readmission risk and analyze the contribution of different lung regions, temporal evolution, and LUS biomarker features. Our results show that lower-lung regions provide the strongest predictive signal and that pleural-line abnormalities are as informative as traditional biomarkers such as A-lines and B-lines. Our best model achieves an F1 score of 0.80, demonstrating the promise of ML-enhanced LUS analysis for practical, noninvasive CHF risk stratification.

\vspace{2em}
\textbf{Related Work.}
Prior work on CHF readmission prediction has relied largely on non-imaging clinical features \cite{Mortazavi2016AnalysisReadmissions,Sharma2022PredictingApproach}, consistently showing the task to be challenging. Lung ultrasound has been extensively studied for diagnostic classification using deep learning \cite{zhang2020contrastive,wang2020contrastive,gare2022weakly}, demonstrating its sensitivity to pulmonary congestion and pleural abnormalities. Recent work \cite{GareLearningTasks} further examined LUS-derived biomarkers, showing them to be predictive across multiple diagnostic tasks. However, the prognostic use of LUS, particularly for forecasting CHF readmission, remains largely unexplored.

\section{Results and Analysis}

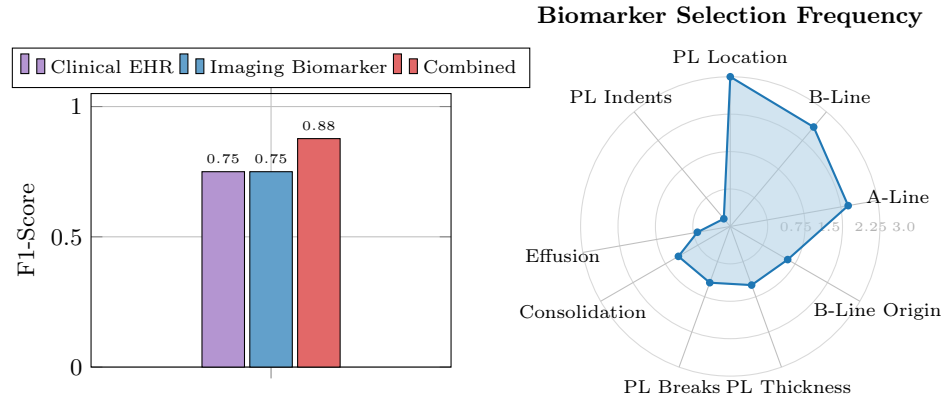
\begin{figure}[t]
\centering

\begin{minipage}[t]{0.52\textwidth}
\centering
\begin{tikzpicture}
\begin{axis}[
    ybar,
    ymin=0.0, ymax=1.05,
    width=\textwidth,
    height=5.2cm,
    bar width=16pt,
    ylabel={F1-Score},
    xtick=data,
    symbolic x coords={F1},
    xticklabels={},
    legend style={
        at={(0.5,1.02)},
        anchor=south,
        font=\scriptsize,
        legend columns=3
    },
    grid=major,
    nodes near coords,
    every node near coord/.append style={font=\tiny},
]

\addplot[fill=clrBaseline!70] coordinates {(F1,0.750)};
\addplot[fill=clrPosterior!70] coordinates {(F1,0.750)};
\addplot[fill=clrAllViews!70] coordinates {(F1,0.877)};

\legend{Clinical EHR, Imaging Biomarker, Combined}

\end{axis}
\end{tikzpicture}
\end{minipage}
\hfill
\begin{minipage}[t]{0.45\textwidth}
\centering
\begin{tikzpicture}[scale=0.9]
  \def\n{9}
  \def\R{2.2}

  \foreach \r/\lab in {0.25/0.75, 0.5/1.5, 0.75/2.25, 1.0/3.0} {
    \draw[gray!30] (0,0) circle (\r*\R);
    \node[font=\tiny, gray!60, anchor=west] at (\r*\R+0.05, 0) {\lab};
  }

  \foreach \ang/\lbl in {
    90/{PL Location},
    50/{B-Line},
    10/{A-Line},
    -30/{B-Line Origin},
    -70/{PL Thickness},
    -110/{PL Breaks},
    -150/{Consolidation},
    -170/{Effusion},
    130/{PL Indents}
  } {
    \draw[gray!50, thin] (0,0) -- (\ang:\R);
    \node[font=\scriptsize,
          inner sep=1pt] at (\ang:\R+0.3) {\lbl};
  }

  \fill[clrPosterior!40, opacity=0.5]
    (90:1.00*\R) --
    (50:0.867*\R) --
    (10:0.800*\R) --
    (-30:0.443*\R) --
    (-70:0.417*\R) --
    (-110:0.400*\R) --
    (-150:0.400*\R) --
    (-170:0.223*\R) --
    (130:0.067*\R) -- cycle;

  \draw[clrPosterior, thick]
    (90:1.00*\R) --
    (50:0.867*\R) --
    (10:0.800*\R) --
    (-30:0.443*\R) --
    (-70:0.417*\R) --
    (-110:0.400*\R) --
    (-150:0.400*\R) --
    (-170:0.223*\R) --
    (130:0.067*\R) -- cycle;

  \foreach \ang/\val in {
    90/1.00, 50/0.867, 10/0.800, -30/0.443,
    -70/0.417, -110/0.400, -150/0.400, -170/0.223, 130/0.067
  } {
    \fill[clrPosterior] (\ang:\val*\R) circle (1.6pt);
  }

  \node[font=\small\bfseries, anchor=south] at (0, \R+0.6)
    {Biomarker Selection Frequency};
\end{tikzpicture}
\end{minipage}

\caption{\textbf{Imaging biomarkers complement clinical EHR.}
\textbf{\textcolor{red}{Left:}} \textbf{EHR vs.\ expert biomarker performance.} On a 25-patient held-out test cohort with available clinical EHR records, the clinical EHR model (purple) and expert LUS biomarker model (blue) achieve identical performance. In contrast, the combined model (red) improves the F1-score by +0.127 and correctly identifies all readmitted patients, highlighting the complementary prognostic information captured by imaging biomarkers that is not present in contemporary clinical EHR data.
\textbf{\textcolor{red}{Right:}} \textbf{Biomarker feature importance.} Each axis represents a LUS biomarker group; radial distance encodes average Random Forest selection frequency (outer ring = 3.0, maximum observed). Pleural Line (PL) Location is the most discriminative single biomarker. The aggregate area covered by pleural attributes (Location, Thickness, Breaks, Indents) rivals the combined A-line and B-line contribution, highlighting underrecognized structural markers as key readmission predictors.
}
\label{fig:combined_side_by_side}
\end{figure}

\section{Method}
\label{sec:method}

We conduct a series of analyses to interpret the key factors driving prognostic predictions. 

\subsection{Dataset}

We pretrained on the lung ultrasound (LUS) dataset from \cite{gare2022learning}, comprising linear-probe B-mode videos acquired on a Sonosite X-Porté system from 221 patients (1001 clips). Each exam includes multiple left and right lung scans at 4–6 cm depth across six standardized views: R1/L1 (upper anterior), R2/L2 (lateral), and R3/L3 (dependent/posterior) (refer Fig~\ref{fig:overview_diagram}).

The main CHF readmission prediction study was conducted on a separately collected CHF-specific cohort acquired using the same imaging protocol and acquisition setup. This cohort consisted of patients admitted with primary congestive heart failure (CHF) and documented 30-day outcomes, yielding 30 patients (9 readmitted, 21 non-readmitted). \textbf{Pilot cohort}: Given the small, imbalanced sample (9:21), results should be interpreted as feasibility evidence rather than definitive clinical validation.

CHF patients were imaged on at least two hospitalization days (some more), yielding 180 scans on Day 1, 150 on Day 2, 66 on Day 3, and 30 on Day 4. Sequential day-pairs were constructed for temporal analysis.

\subsection{Pretrained Feature Extraction}
A Temporal Shift Module (TSM) network \cite{lin2019tsm} with a ResNet-18 backbone was pretrained to regress the S/F ratio from LUS clips. The encoder was then frozen, and 512-dimensional embeddings were extracted from the layer preceding the regression head to provide high-level spatiotemporal features. For each patient, embeddings were generated at all available time points, and sequential day-pairs (e.g., Day 1–2) were formed for downstream readmission classification using each patient’s single readmission label. Since most patients had only Day 1 and Day 2 scans, our primary model was trained on Day 1–Day 2 pairs, with an All-Days variant trained on all available pairs (Table \ref{tab:view_day_comparison}). Each LUS exam included six views (R1–L3), and models were trained separately for each view to assess per-view predictive value, with additional experiments evaluating multi-view combinations.





\subsection{Nested Cross-Validation and Evaluation}
We used five-by-five nested cross-validation to obtain unbiased performance estimates. Patients were split into five outer folds (six patients each) with strict patient-level separation to prevent leakage. Hyperparameter tuning was performed exclusively in the inner loop \cite{varma}, while outer folds were reserved for testing.
In each outer iteration, one fold served as test data and the remaining four formed the inner loop (three training, one validation, rotated). The model achieving the highest weighted F1 on inner validation was selected.
This procedure yielded fold-specific results for all classifiers and fusion setups. For multi-view or multi-day experiments, decision-level fusion aggregated predictions across clips and time points.
\textbf{Evaluation:} Final performance was computed from pooled outer-fold predictions using weighted F1-score. Ninety-five percent confidence intervals were estimated via bootstrap resampling (2000 iterations, 2.5th–97.5th percentiles).

\subsection{View-Specific Performance \& Classifier Comparison}
Table~\ref{tab:view_results} summarizes F1-scores (with 95\% CIs) for each view and classifier on Day~1–Day~2 pairs using the temporal \textit{difference} representation and no cross-lung pooling. Lower posterior views (Left-3 \& Right-3) consistently show higher predictive performance than anterior views, supporting the intuition that dependent regions better capture pulmonary congestion. Across classifiers, MLP and MLP-Large generally achieve the best F1-scores, particularly when all views are concatenated.

\begin{table*}[ht]
\centering
\small
\resizebox{\textwidth}{!}
{\begin{tabular}{lcccccc}
\hline
View & Decision-Tree & Random-Forest & SVM & MLP & MLP-Large & TabPFN \\
\hline
Left-1   & 0.60 [0.40–0.80]          & 0.55 [0.34–0.77]          & 0.41 [0.21–0.62]          & 0.53 [0.33–0.73]          & 0.53 [0.33–0.73]          & \textbf{0.53 [0.31–0.77]} \\
Left-2   & 0.42 [0.21–0.63]          & 0.55 [0.34–0.74]          & 0.49 [0.29–0.68]          & 0.57 [0.37–0.76]          & 0.49 [0.29–0.68]          & \textbf{0.53 [0.32–0.74]} \\
Left-3   & \textbf{0.63 [0.43–0.82]} & \textbf{0.58 [0.36–0.80]} & \textbf{0.60 [0.39–0.78]} & \textbf{0.68 [0.48–0.84]} & \textbf{0.68 [0.48–0.84]} & 0.50 [0.28–0.72] \\
\hline
Right-1   & 0.41 [0.21–0.60]          & \textbf{0.64 [0.43–0.84]} & \textbf{0.65 [0.45–0.83]} & 0.44 [0.24–0.65]          & 0.57 [0.37–0.76]          & 0.50 [0.29–0.71] \\
Right-2   & 0.41 [0.22–0.62]          & \textbf{0.64 [0.40–0.86]} & 0.52 [0.32–0.73]          & 0.47 [0.25–0.68]          & 0.47 [0.25–0.69]          & 0.51 [0.29–0.75] \\
Right-3   & \textbf{0.50 [0.29–0.69]} & 0.55 [0.34–0.77]          & 0.55 [0.34–0.75]          & \textbf{0.61 [0.40–0.82]} & \textbf{0.68 [0.48–0.86]} & \textbf{0.64 [0.43–0.84]} \\
\hline
All Views & 0.53 [0.32–0.73] & \textbf{0.64 [0.40–0.85]} & 0.58 [0.36–0.80] & \textbf{0.80 [0.62–0.96]} & \textbf{0.75 [0.55–0.92]} & 0.55 [0.34–0.78] \\
\hline
\end{tabular}}
\caption{\textbf{View-wise classifier comparison.} F1-scores (95\% CIs) for each lung ultrasound (LUS) view on Day~1 vs.~Day~2 using the \textit{difference} representation and not trained with left–right pooled data. The ``All Views'' configuration combines all six views by concatenating their embeddings. MLP and MLP-Large perform best overall, with posterior views (3 and 6) most predictive.}
\label{tab:view_results}
\end{table*}

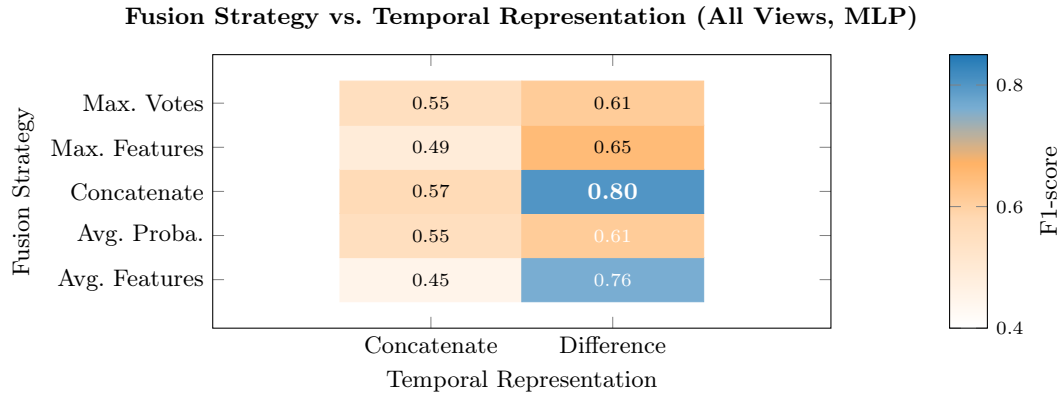
\begin{figure}[!ht]
\centering
\begin{tikzpicture}
\begin{axis}[
    colormap={coolwarm}{
        color(0)=(white)
        color(0.4)=(orange!30)
        color(0.6)=(orange!60)
        color(0.8)=(clrPosterior!60)
        color(1.0)=(clrPosterior)
    },
    colorbar,
    colorbar style={
        ylabel={F1-score},
        ylabel style={font=\small},
        yticklabel style={font=\scriptsize},
    },
    point meta min=0.40,
    point meta max=0.85,
    width=0.80\columnwidth,
    height=5.2cm,
    xlabel={Temporal Representation},
    ylabel={Fusion Strategy},
    xlabel style={font=\small},
    ylabel style={font=\small},
    xtick={0,1},
    xticklabels={Concatenate, Difference},
    xticklabel style={font=\small},
    ytick={0,1,2,3,4},
    yticklabels={
        {Avg.\ Features},
        {Avg.\ Proba.},
        {Concatenate},
        {Max.\ Features},
        {Max.\ Votes}
    },
    yticklabel style={font=\small},
    title={\small \textbf{Fusion Strategy vs.\ Temporal Representation (All Views, MLP)}},
    title style={font=\small},
    enlarge x limits=0.35,
    enlarge y limits=0.12,
]
\addplot[
    matrix plot*,
    point meta=explicit,
    mesh/rows=5,
    mesh/cols=2,
] table[meta=f1] {
    x y f1
    0 0 0.45
    1 0 0.76
    0 1 0.55
    1 1 0.61
    0 2 0.57
    1 2 0.80
    0 3 0.49
    1 3 0.65
    0 4 0.55
    1 4 0.61
};
\node[font=\scriptsize] at (axis cs:0,0) {0.45};
\node[font=\scriptsize,white] at (axis cs:1,0) {0.76};
\node[font=\scriptsize]       at (axis cs:0,1) {0.55};
\node[font=\scriptsize,white] at (axis cs:1,1) {0.61};
\node[font=\scriptsize]       at (axis cs:0,2) {0.57};
\node[font=\scriptsize,white,font=\small\bfseries] at (axis cs:1,2) {0.80};
\node[font=\scriptsize]       at (axis cs:0,3) {0.49};
\node[font=\scriptsize]       at (axis cs:1,3) {0.65};
\node[font=\scriptsize]       at (axis cs:0,4) {0.55};
\node[font=\scriptsize]       at (axis cs:1,4) {0.61};
\end{axis}
\end{tikzpicture}
\caption{\textbf{Multi-view fusion heatmap (MLP, All Views).} F1-scores across five fusion strategies (rows) and two temporal representations (columns). Feature concatenation with temporal difference achieves the best performance (\textbf{0.80}). Consistently higher scores in the ``Difference'' column indicate that temporal representation has greater impact than fusion strategy.}
\label{tab:mlp_fusion}
\end{figure}

\vspace{-1.5em}
\subsection{Multi-View Fusion}



To leverage spatial information, we evaluated two fusion strategies.
\textit{Feature-level fusion} combines embeddings from all six views via averaging, max pooling, or concatenation.
\textit{Decision-level fusion} aggregates per-view predictions by either max-voting or averaging the predicted probabilities across views.
As shown in Fig.~\ref{tab:mlp_fusion}, feature concatenation paired with temporal difference yields the best performance.


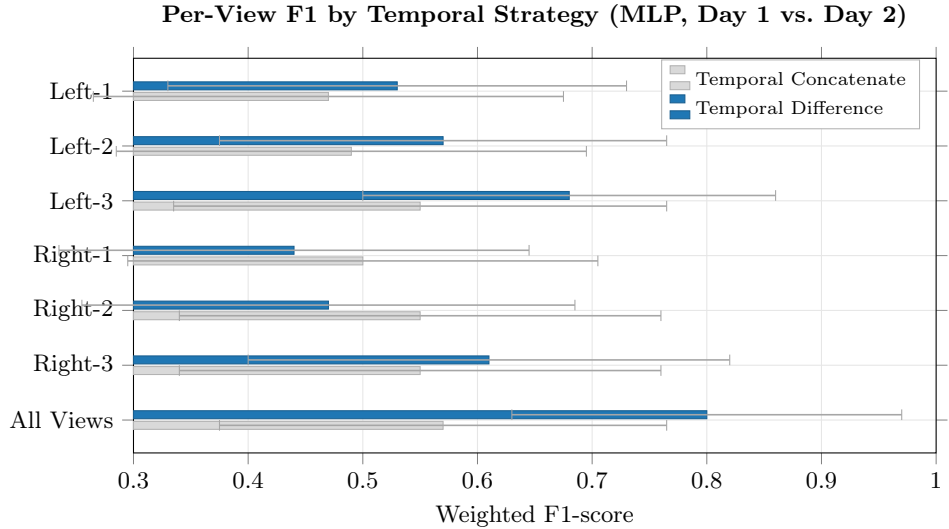
\begin{figure*}[!t]
\centering
\begin{tikzpicture}
\begin{axis}[
    xbar,
    xmin=0.30, xmax=1.00,
    width=\columnwidth,
    height=6.8cm,
    bar width=3pt,
    xlabel={Weighted F1-score},
    xlabel style={font=\small},
    symbolic y coords={
        {All Views},
        {Right-3},
        {Right-2},
        {Right-1},
        {Left-3},
        {Left-2},
        {Left-1}
    },
    ytick=data,
    yticklabel style={font=\small},
    legend style={
        at={(0.98,0.82)},
        anchor=south east,
        font=\scriptsize,
        legend cell align=left,
        fill=white,
        draw=gray!60
    },
    legend columns=1,
    tick align=outside,
    grid=major,
    grid style={gray!20},
    clip=false,
    title={\small \textbf{Per-View F1 by Temporal Strategy (MLP, Day 1 vs.\ Day 2)}},
    title style={font=\small, yshift=2pt},
    error bars/y dir=none,
    error bars/x dir=both,
    error bars/x explicit,
    error bars/error bar style={line width=0.6pt, gray!70},
]

\addplot[
    xbar,
    fill=gray!30,
    draw=gray!60,
    bar shift=-2.pt,
    error bars/.cd, x dir=both, x explicit
] coordinates {
    (0.47,{Left-1})   +- (0.205,0)
    (0.49,{Left-2})   +- (0.205,0)
    (0.55,{Left-3})   +- (0.215,0)
    (0.50,{Right-1})  +- (0.205,0)
    (0.55,{Right-2})  +- (0.210,0)
    (0.55,{Right-3})  +- (0.210,0)
    (0.57,{All Views})+- (0.195,0)
};

\addplot[
    xbar,
    fill=clrPosterior,
    draw=clrPosterior!80!black,
    bar shift=2.pt,
    error bars/.cd, x dir=both, x explicit
] coordinates {
    (0.53,{Left-1})   +- (0.200,0)
    (0.57,{Left-2})   +- (0.195,0)
    (0.68,{Left-3})   +- (0.180,0)
    (0.44,{Right-1})  +- (0.205,0)
    (0.47,{Right-2})  +- (0.215,0)
    (0.61,{Right-3})  +- (0.210,0)
    (0.80,{All Views})+- (0.170,0)
};


\legend{Temporal Concatenate, Temporal Difference}
\end{axis}
\end{tikzpicture}
\caption{
Weighted F1 (95\% CI) for temporal concatenation (gray) vs.\ temporal difference (blue) across six LUS views and the multi-view setting. Posterior views (Left-3, Right-3) outperform anterior views; temporal difference yields consistent gains; and all-view fusion achieves the highest F1.}

\label{tab:mlp_mode1_comparison}
\end{figure*}

\subsection{Temporal Fusion}
To model temporal progression, we compared two ways of combining sequential embeddings: (1) taking their difference (e.g., Day 2–Day 1) to highlight changes, and (2) concatenating them to preserve both time points.

Fig.~\ref{tab:mlp_mode1_comparison} shows that using the difference between sequential scans generally outperforms concatenation, particularly for the posterior views and the multi-view setting. This supports the hypothesis that capturing progression or improvement over time is critical for readmission risk.


\subsection{Day-Pair Modeling}
We next evaluated whether training on all available day-pair combinations improves generalization. Table~\ref{tab:view_day_comparison} compares three settings: (1) a model trained and tested on Day 1 – Day 2 pairs only (Single-Day model), (2) a model trained on all day-pairs but evaluated on Day~1–Day~2 pairs (All-Days model), and (3) the All-Days model evaluated on all day-pairs.

Overall, the Single-Day model yields the highest Day~1 –Day~2 F1 (0.80), while the All-Days model offers more consistent performance across views and day-pairs, indicating better temporal robustness.

\begin{table*}[ht]
\centering
\small
\setlength{\tabcolsep}{6pt}
\renewcommand{\arraystretch}{1.1}
\resizebox{\textwidth}{1.5cm}{
\begin{tabular}{lccc}
\hline
View &
\textbf{Day 1 vs Day 2 (Single-Day Model)} &
\textbf{Day 1 vs Day 2 (All-Days Model)} &
\textbf{All Days (All-Days Model)} \\
\hline
Left-1 & 0.53 [0.33–0.73] & 0.63 [0.42–0.82] & 0.66 [0.52–0.79] \\
Left-2 & 0.57 [0.37–0.76] & 0.60 [0.40–0.79] & 0.62 [0.47–0.75] \\
Left-3 & 0.68 [0.48–0.84] & 0.44 [0.23–0.67] & 0.48 [0.33–0.63] \\
\hline
Right-1 & 0.44 [0.24–0.65] & 0.65 [0.46–0.83] & 0.64 [0.50–0.77] \\
Right-2 & 0.47 [0.25–0.68] & 0.38 [0.19–0.57] & 0.46 [0.32–0.59] \\
Right-3 & 0.61 [0.40–0.82] & 0.68 [0.49–0.85] & 0.65 [0.52–0.78] \\
\hline
All Views & 0.80 [0.62–0.96] & 0.72 [0.53–0.88] & 0.72 [0.60–0.83] \\
\hline
\end{tabular}
}
\caption{\textbf{Day-pair modeling.} F1-scores (95\% CI) for each LUS view under different training settings. ``Day~1 vs.~Day~2 (Single-Day Model)'' refers to performance on Day~1–Day~2 pairs using a model trained only on those pairs. ``Day~1 vs.~Day~2 (All-Days Model)'' uses a model trained on all available day-pair combinations but evaluated on Day~1–Day~2. ``All Days (All-Days Model)'' reports performance on all day-pairs. Fixed conditions: temporal \textit{difference}, no cross-lung pooling; for ``All Views'', embeddings from all views are concatenated.}
\label{tab:view_day_comparison}
\end{table*}

\vspace{-1.5em}
\subsection{Cross-Lung: Training with Left–Right Pooled Data}

We exploited approximate left–right lung symmetry by treating paired views (R1–L1, R2–L2, R3–L3) as equivalent. During training, all directional mappings (e.g., R1→L1, L1→R1) were included to augment the data and enforce cross-lung consistency. This pooling was applied only in training; evaluation remained side-specific to preserve clinical interpretability.

\begin{figure*}[!t]
\centering
\begin{tikzpicture}
\begin{axis}[
    ybar,
    ymin=0.30, ymax=1.00,
    width=\columnwidth,
    height=4.5cm,
    bar width=8pt,
    ylabel={Weighted F1-score},
    ylabel style={font=\small},
    xtick=data,
    symbolic x coords={Left-1, Left-2, Left-3, Right-1, Right-2, Right-3, {All Views}},
    xticklabel style={font=\tiny, rotate=30, anchor=east},
    legend style={
        at={(0.02,0.98)},
        anchor=north west,
        font=\scriptsize,
        fill=white,
        draw=gray!60
    },
    legend columns=1,
    grid=major,
    grid style={gray!20},
    clip=false,
    title={\small \textbf{Effect of Cross-Lung Augmentation (MLP, Temporal Difference)}},
    title style={font=\small},
    error bars/y dir=both,
    error bars/y explicit,
    error bars/error bar style={line width=0.6pt, gray!60},
]
\addplot[
    ybar,
    fill=gray!30,
    draw=gray!60,
    bar shift=-5pt,
    error bars/.cd, y dir=both, y explicit
] coordinates {
    (Left-1,  0.53) +- (0,0.200)
    (Left-2,  0.57) +- (0,0.195)
    (Left-3,  0.68) +- (0,0.180)
    (Right-1, 0.44) +- (0,0.205)
    (Right-2, 0.47) +- (0,0.215)
    (Right-3, 0.61) +- (0,0.210)
    ({All Views}, 0.80) +- (0,0.170)
};
\addplot[
    ybar,
    fill=clrPosterior,
    draw=clrPosterior!80!black,
    bar shift=5pt,
    error bars/.cd, y dir=both, y explicit
] coordinates {
    (Left-1,  0.63) +- (0,0.200)
    (Left-2,  0.57) +- (0,0.190)
    (Left-3,  0.75) +- (0,0.185)
    (Right-1, 0.50) +- (0,0.215)
    (Right-2, 0.61) +- (0,0.210)
    (Right-3, 0.69) +- (0,0.185)
    ({All Views}, 0.80) +- (0,0.170)
};


\legend{No Cross-Lung, With Cross-Lung}
\end{axis}
\end{tikzpicture}
\caption{\textbf{Cross-lung training.} Comparison of MLP performance with and without left–right pooled augmentation across all LUS views. Fixed conditions: Day~1 vs.~Day~2, temporal \textit{difference}; for ``All Views'', embeddings are concatenated across views. Cross-lung pooling improves or maintains performance for most views while preserving the best multi-view result.}
\label{tab:mlp_crosslung_comparison}
\end{figure*}
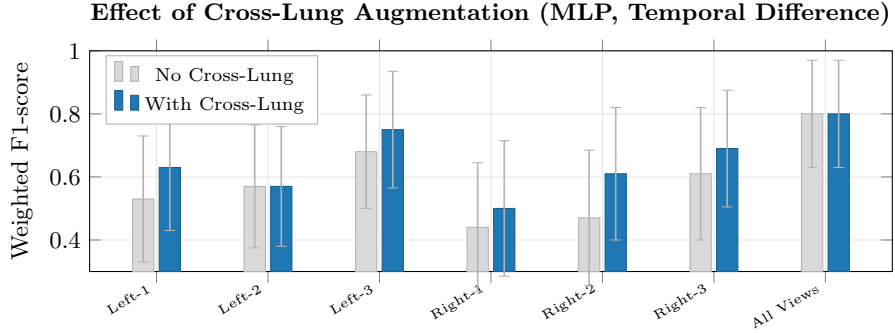

As shown in Fig.~\ref{tab:mlp_crosslung_comparison}, cross-lung pooling improves or maintains performance for most individual views (notably L3 and R3) while preserving the multi-view result (All Views F1 = 0.80). Thus, it serves as effective augmentation without degrading the best-performing configuration.

\subsection{Biomarker-Prediction Pretraining}

We also evaluated a pretrained network from \cite{gare2022learning} designed to regress 38 interpretable LUS biomarkers, including A-lines, B-lines, pleural abnormalities, consolidations, and effusions. Using the same backbone pretraining dataset as the S/F model, the resulting 38-dimensional biomarker prediction vector was used in place of the 512-dimensional TSM embedding and evaluated using the same downstream classifiers and cross-validation protocol described in Sec.~\ref{sec:method}. This analysis aimed to assess the predictive utility of clinically interpretable biomarker-level features for 30-day readmission.

\begin{table*}[ht]
\centering
\small
\resizebox{\textwidth}{!}{
\begin{tabular}{lcccccc}
\hline
View & \textbf{Decision-Tree} & \textbf{Random-Forest} & \textbf{SVM} & \textbf{MLP} & \textbf{MLP-Large} & \textbf{TabPFN} \\
\hline
Left-3 (Day 1 vs. Day 2) &
0.63 [0.42–0.84] &
0.55 [0.33–0.77] &
0.44 [0.23–0.65] &
0.50 [0.28–0.71] &
0.53 [0.31–0.74] &
\textbf{0.58 [0.36–0.78]} \\

Left-3 (All Days) &
0.52 [0.38–0.66] &
0.56 [0.41–0.70] &
0.47 [0.34–0.62] &
0.55 [0.39–0.68] &
0.55 [0.40–0.69] &
\textbf{0.58 [0.42–0.72]} \\

\hline
Right-3 (Day 1 vs. Day 2) &
0.45 [0.25–0.66] &
0.50 [0.29–0.71] &
0.50 [0.29–0.69] &
0.47 [0.26–0.68] &
0.47 [0.26–0.69] &
0.39 [0.20–0.62] \\

Right-3 (All Days) &
0.49 [0.35–0.62] &
0.53 [0.38–0.68] &
0.55 [0.42–0.68] &
0.50 [0.36–0.63] &
0.58 [0.43–0.71] &
0.55 [0.38–0.70] \\

\hline
All Views (Day 1 vs. Day 2) &
0.58 [0.37–0.76] &
0.53 [0.31–0.77] &
\textbf{0.71 [0.47–0.90]} &
0.54 [0.34–0.73] &
\textbf{0.60 [0.39–0.80]} &
0.49 [0.27–0.71] \\

All Views (All Days) &
\textbf{0.68 [0.54–0.80]} &
\textbf{0.67 [0.52–0.80]} &
0.67 [0.52–0.80] &
\textbf{0.65 [0.52–0.78]} &
\textbf{0.60 [0.46–0.74]} &
0.53 [0.37–0.69] \\
\hline
\end{tabular}}
\caption{\textbf{Biomarker-based pretraining.} Using predicted biomarker values as input features, F1-scores (95\% CI) with temporal \textit{difference} are reported for the top performing views (L3, R3) and the multi-view condition, for Day~1 vs.~Day~2 and All-Days models. Fixed conditions: no cross-lung pooling; the ``All Views'' configuration combines all six views by feature concatenation.}
\label{tab:biomarker_res_2}
\end{table*}

Table~\ref{tab:biomarker_res_2} summarizes results obtained using the 38-dimensional biomarker features for the top-performing individual views and multi-view fusion strategies. The best-performing biomarker-based model was the All Views (Day 1 vs.\ Day 2) SVM configuration, achieving an F1-score of 0.71, which remained below the best-performing 512-dimensional TSM embedding model (F1-score = 0.80).

Fig.~\ref{fig:combined_side_by_side} (right) presents a radar plot in which each axis corresponds to a LUS biomarker group, while the radial distance represents the average Random Forest feature-selection frequency across folds. A-line, B-line, and pleural line features (including location, thickness, and continuity) were selected most consistently, indicating that markers related to lung aeration and pleural structure are the strongest predictors of readmission risk. The Random Forest feature rankings further emphasize the combined importance of A-line, B-line, and pleural descriptors, consistent with clinical intuition.

\begin{figure}[t]
    \centering
    \subfloat[Readmitted\label{fig:gradcam_readmitted}]{
    \includegraphics[width=0.45\linewidth]{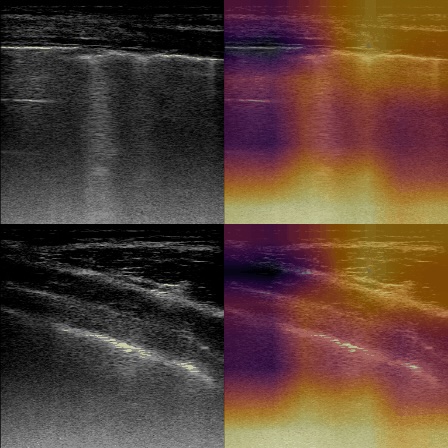}
    }
    \hfill
    \subfloat[Not Readmitted\label{fig:gradcam_notreadmitted}]{
        \includegraphics[width=0.45\linewidth]{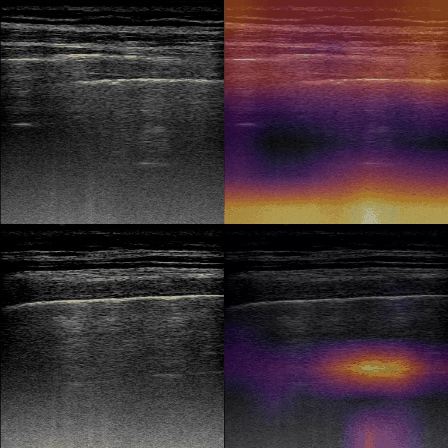}
    }
    \caption{\textbf{Grad-CAM visualizations} for paired Day~1 (top row) and Day~2 (bottom row) LUS clips. (a) Readmitted patient showing marked Day~2 effusion and corresponding focused model attention. (b) Not-readmitted patient demonstrating stable Day~2 appearance with diffuse, low-intensity attention.}
    \label{fig:gradcam_res}
\end{figure}

\subsection{Interpretability via Grad-CAM}
We applied Grad-CAM \cite{SelvarajuGrad-CAM:Localization} to the final convolutional block of the pretrained TSM–ResNet18 encoder to visualize regions driving readmission predictions. Since the model processes paired sequential exams (e.g., Day 1 and 2), Grad-CAM was generated for both time points under the same prediction, yielding paired visual explanations. Analyses were limited to outer-fold test patients. Figure~\ref{fig:gradcam_res} shows representative readmitted and non-readmitted cases.

\section{Conclusion}
\label{sec:conclusion}
This work demonstrates that lung ultrasound provides informative prognostic signals for 30-day CHF readmission. ML models achieved strong performance (F1 up to 0.80), driven largely by lower-lung and pleural-line abnormalities. Temporal difference features and multi-view concatenation were most effective, with cross-lung augmentation offering additional gains. Biomarker models highlighted A-lines, B-lines, and pleural descriptors as key predictors, underscoring the value of LUS for noninvasive CHF risk stratification.

\vspace{2em}
\textbf{Acknowledgments.}
This research was partially supported by the Liang Zhao Endowed Fellowship at Carnegie Mellon University (CMU). The authors gratefully acknowledge the clinicians at Louisiana State University (LSU) for their assistance with data collection, as well as collaborators at CMU for their helpful feedback and discussions. All datasets were de-identified prior to analysis. The LSU dataset was collected under IRB protocol \#1509, titled \textit{Artificial Intelligence Interpretation of Lung Ultrasound Images} and was de-identified prior to transfer to CMU.

\vfill
\pagebreak


\bibliographystyle{splncs04}
\bibliography{strings,refs,short_ref,references}

@article{Mortazavi2016AnalysisReadmissions,
    title = {{Analysis of Machine Learning Techniques for Heart Failure Readmissions}},
    year = {2016},
    journal = {Circulation. Cardiovascular quality and outcomes},
    author = {Mortazavi, Bobak J. and Downing, Nicholas S. and Bucholz, Emily M. and Dharmarajan, Kumar and Manhapra, Ajay and Li, Shu Xia and Negahban, Sahand N. and Krumholz, Harlan M.},
    number = {6},
    month = {11},
    pages = {629--640},
    volume = {9},
    publisher = {Circ Cardiovasc Qual Outcomes},
    url = {https://pubmed.ncbi.nlm.nih.gov/28263938/},
    doi = {10.1161/CIRCOUTCOMES.116.003039},
    issn = {1941-7705},
    pmid = {28263938}
}

@article{SelvarajuGrad-CAM:Localization,
    title = {{Grad-CAM: Visual Explanations from Deep Networks via Gradient-based Localization}},
    author = {Selvaraju, Ramprasaath R and Cogswell, Michael and Das, Abhishek and Vedantam, Ramakrishna and Parikh, Devi and Batra, Dhruv},
    url = {https://github.com/},
    arxivId = {1610.02391v4}
}

@article{GareLearningTasks,
    title = {{Learning Generic Lung Ultrasound Biomarkers for Decoupling Feature Extraction from Downstream Tasks}},
    author = {Gare, Gautam Rajendrakumar and Fox, Tom and Lowery, Pete and Zamora, Kevin and Tran, Hai V and Hutchins, Laura and Montgomery, David and Krishnan, Amita and Ramanan, Deva Kannan and Luis Rodriguez, Ricardo and Deboisblanc, Bennett P and Galeotti, John Michael},
    isbn = {2206.08398v1},
    arxivId = {2206.08398v1}
}

@article{Harvey2021P004.Failure,
    title = {{P004. 30-Day Hospital Readmission Rates, Frequencies, and Heart Failure Classification for Patients with Heart Failure}},
    year = {2021},
    journal = {Heart {\&} Lung},
    author = {Harvey, Margaret and Brown, Yolanda and Goyer, Twonia},
    number = {4},
    month = {7},
    pages = {562},
    volume = {50},
    publisher = {Mosby},
    url = {https://www.sciencedirect.com/science/article/pii/S0147956321001163},
    doi = {10.1016/J.HRTLNG.2021.03.059},
    issn = {0147-9563}
}

@article{Sharma2022PredictingApproach,
    title = {{Predicting 30-Day Readmissions in Patients With Heart Failure Using Administrative Data: A Machine Learning Approach}},
    year = {2022},
    journal = {Journal of Cardiac Failure},
    author = {Sharma, VISHAL and KULKARNI, VINAYKUMAR and MCALISTER, FINLAY and EURICH, D. E.A.N. and KESHWANI, SHANIL and SIMPSON, SCOT H. and VOAKLANDER, D. O.N. and SAMANANI, SALIM},
    number = {5},
    month = {5},
    pages = {710--722},
    volume = {28},
    publisher = {Elsevier B.V.},
    url = {https://pubmed.ncbi.nlm.nih.gov/34936894/},
    doi = {10.1016/j.cardfail.2021.12.004},
    issn = {15328414},
    pmid = {34936894}
}

@article{varma,
author = {Varma, Sudhir and Simon, Richard},
year = {2006},
month = {02},
pages = {91},
title = {Bias in Error Estimation When Using Cross-Validation for Model Selection.” BMC Bioinformatics, 7(1), 91},
volume = {7},
journal = {BMC bioinformatics},
doi = {10.1186/1471-2105-7-91}
}

@inproceedings{lin2019tsm,
  title={Tsm: Temporal shift module for efficient video understanding},
  author={Lin, Ji and Gan, Chuang and Han, Song},
  booktitle={Proceedings of the IEEE/CVF International Conference on Computer Vision},
  pages={7083--7093},
  year={2019}
}

@article{gare2022learning,
  title={Learning Generic Lung Ultrasound Biomarkers for Decoupling Feature Extraction from Downstream Tasks},
  author={Gare, Gautam Rajendrakumar and Fox, Tom and Lowery, Pete and Zamora, Kevin and Tran, Hai V and Hutchins, Laura and Montgomery, David and Krishnan, Amita and Ramanan, Deva Kannan and Rodriguez, Ricardo Luis and others},
  journal={arXiv preprint arXiv:2206.08398},
  year={2022}
}

@article{gare2022weakly,
  title={Weakly Supervised Contrastive Learning for Better Severity Scoring of Lung Ultrasound},
  author={Gare, Gautam Rajendrakumar and Tran, Hai V and deBoisblanc, Bennett P and Rodriguez, Ricardo Luis and Galeotti, John Michael},
  journal={arXiv preprint arXiv:2201.07357},
  year={2022}
}

@article{zhang2020contrastive,
  title={Contrastive learning of medical visual representations from paired images and text},
  author={Zhang, Yuhao and Jiang, Hang and Miura, Yasuhide and Manning, Christopher D and Langlotz, Curtis P},
  journal={arXiv preprint arXiv:2010.00747},
  year={2020}
}

@article{wang2020contrastive,
  title={Contrastive cross-site learning with redesigned net for COVID-19 CT classification},
  author={Wang, Zhao and Liu, Quande and Dou, Qi},
  journal={IEEE Journal of Biomedical and Health Informatics},
  volume={24},
  number={10},
  pages={2806--2813},
  year={2020},
  publisher={IEEE}
}

\end{document}